\documentclass[aps,preprint,onecolumn,floatfix,nofootinbib,superscriptaddress]{revtex4-1}

\usepackage{amsmath}
\usepackage{amssymb}
\usepackage{amsfonts}
\usepackage{float}
\usepackage{mathtools}                                  % complete mathematical tools
\usepackage{listings}                                   % put code snippet into text
\usepackage{bm}                                         % bold math
\usepackage{subfig}                                     % put more figures together

\usepackage{color}

\DeclareMathOperator{\Tr}{Tr}

\DeclareMathOperator{\adj}{adj}

\newcommand{\rom}[1]{\uppercase\expandafter{\romannumeral #1\relax}}    % Roman numbers 

\makeatletter
\def\l@subsection#1#2{}
\def\l@subsubsection#1#2{}
\makeatother

\def\npart{N_\text{p}}
\begin{document}

\pagenumbering{arabic}

% \title{Generalized constraints for the rigorous elimination of the global translation in explicitly correlated Gaussian functions}
\title{Generalized elimination of the global translation from explicitly correlated Gaussian functions}

\author{Andrea Muolo}
\affiliation{ETH Z\"urich, Laboratory of Physical Chemistry, Vladimir-Prelog-Weg 2, 8093 Z\"urich, Switzerland}

\author{Edit M\'atyus}
\email{Corresponding author: matyus@chem.elte.hu}
\affiliation{Institute of Chemistry, E\"otv\"os Lor\'and University, P\'azm\'any P\'eter s\'et\'any 1/A, 1117 Budapest, Hungary}

\author{Markus Reiher}
\email{Corresponding author: markus.reiher@phys.chem.ethz.ch}
\affiliation{ETH Z\"urich, Laboratory of Physical Chemistry, Vladimir-Prelog-Weg 2, 8093 Z\"urich, Switzerland}

\begin{abstract}
\noindent This paper presents the multi-channel generalization of 
the center-of-mass kinetic energy elimination approach [Mol. Phys., 111 2086 (2013)]
when the Schr\"odinger equation is solved variationally with explicitly correlated Gaussian functions.
The approach has immediate relevance in many-particle systems which are handled without 
the Born--Oppenheimer approximation and can be employed also for Dirac-type Hamiltonians. 
The practical realization and numerical properties of solving the Schr\"odinger equation in 
laboratory-frame Cartesian coordinates
are demonstrated for the ground rovibronic state
of the H$_2^+=\lbrace\text{p}^+,\text{p}^+,\text{e}^+\rbrace$ ion and 
the H$_2=\lbrace\text{p}^+,\text{p}^+,\text{e}^+,\text{e}^+\rbrace$ molecule.

\end{abstract}

\maketitle

\section{Introduction}\label{SEC:intro}
\noindent%
Explicitly correlated Gaussian (\texttt{ECG}) functions have a long history in variational calculations of 
few-particle quantum mechanical systems \cite{ECG-history1960,ECG-history1960_1,ECG-history1977,ECG-history1986,
ECG-history1993,Adamowicz2013,Matyus2012} yielding 
results with a nano-Hartree accuracy in the energy.
An obvious advantage of these functions is that analytic matrix elements can be derived for 
almost all physically relevant operators and for an arbitrary number of particles.
This general applicability is a particularly important advantage for our work, 
in which we aim to develop a general approach for atoms, molecules, 
or other more exotic molecular ``objects'' (e.g.,
positronium complexes) 
by considering all particles on equal footing, i.e., 
without introducing the Born--Oppenheimer (BO) approximation.
We refer to this framework as a pre-Born--Oppenheimer (pre-BO) theory 
in order to emphasize the departure from the traditional 
(and undoubtedly very successful) Born--Oppenheimer approximation
and other ``post-Born--Oppenheimer'' approaches, 
which correct for or go beyond the BO approximation.

The spatial symmetry properties of the pre-BO approach are reminiscent of nuclear motion theory 
(or also called ``quantum dynamics'')
in which the rovibrational Schr\"odinger equation is solved on some potential energy surface. In both cases, 
the full Hamiltonian has a continuous spectrum due to the overall translation of the system. 
In order to obtain the translation-free, i.e., 
translationally invariant, properties, the laboratory-frame Cartesian coordinates (LFCC) are 
traditionally replaced by a translationally invariant set of Cartesian coordinates (TICC)
and the Cartesian coordinates of the center of mass (CMCC). 
This linear transformation \cite{CSTSutcliffe}---%
although rather simple in comparison to the commonly introduced body-fixed frame, orientational angles, and 
curvilinear internal coordinates to efficiently describe rotating-vibrating 
molecular systems \cite{rovib1,rovib2,rovib3,rovib4}---makes the original, 
very simple Cartesian kinetic energy operator more complicated. 
Certainly, the resulting TICC kinetic energy operator (after subtracting the 
center-of-mass kinetic energy term) has been successfully used many times, 
see for example \cite{Adamowicz2003a,Matyus2011a,Matyus2011b,Matyus2012,Adamowicz2013,Matyus2013}, 
it is also reasonable to ask whether it is possible to avoid any coordinate change at all and stay with the original,
mathematically and conceptually very simple laboratory-fixed Cartesian coordinates. 
One might ask why to look for an alternative to the already working TICC approach---we ask: why not?
To give a historical example in which similar questions resulted in important developments, we mention 
the numerical evaluation of the diagonal Born--Oppenheimer correction (DBOC) (within the post-BO framework).
The calculation of the DBOC
has been made extremely elegant and simple by Handy and co-workers \cite{HaLe86,HaYaSc86,CeKu97,Ku97} by using 
laboratory-fixed Cartesian coordinates, instead
of the earlier used more tedious way of \text{choosing} some TICC set and transforming the relevant expressions
to this TICC and CMCC coordinate set.

Back to our pre-BO framework, Ref.~\cite{Benjamin2013} has shown that it is possible to calculate
the translation-free part of the spectrum, i.e., rotational-vibrational-electronic levels, 
of any isolated many-particle system by solving the Schr\"odinger equation in laboratory-frame Cartesian
coordinates. In this approach, 
the CM kinetic energy contribution is cancelled during the integral evaluation of
the ECG basis functions.

Although the integral evaluation with ECGs is straightforward, 
their parameterization---which is after all a very high-dimensional parameterization problem---requires special care in
particular when highly accurate energy levels of molecular systems (i.e., assemblies of light and heavy particles) 
are to be calculated. The LFCC approach of
Ref.~\cite{Benjamin2013} was developed for a certain way of parameterization. The present work 
generalizes this LFCC approach and makes it applicable together with the most general ``multi-channel optimization''
of the ECG parameter set, in which the optimization approach cycles through 
various pairs of particles and groups of particles and varies the ECG parameters (exponents)
to describe the interaction of these pairs or groups optimally.

To this end, we had to study in detail 
the general properties of the LF$\rightarrow$(TI,CM) Cartesian coordinate transformation, as well as
the analytic kinetic energy expectation value expressions, 
which is described in the first part of the article. 
In the second part, we demonstrate the general applicability of this generalized LFCC approach and 
the excellent numerical and convergence properties of the multi-channel optimization.
As ``relativistic effects'' have been shown to be equally important to 
``nonadiabatic effects'' in light systems, see for example Ref.~\cite{PiJe09}, we emphasize 
that the LFCC approach developed in the present paper
is transferable to the Dirac theory which we will consider in future work.

\section{The Schr\"odinger Hamiltonian and coordinate sets}\label{SEC:theory}
\noindent%
Given a collection of Cartesian coordinates 
let us consider the laboratory-frame (LF) Cartesian coordinates, 
$\bm{r}=\left(\bm{r}_1,\ldots,\bm{r}_{\npart}\right)^T$, of
$\npart$\ particles associated with some $m_i$ masses and 
$q_i$ electric charges, which parameterize the instantaneous Coulomb interactions acting among the 
particles. The Schr\"odinger Hamiltonian, in Hartree atomic units %($m_{e}=1$, $e=1$, $\hbar=1$, and $4\pi\epsilon_0=1$) 
is 
\begin{equation}
\label{nonrel-H}
\hat{H}_{\text{S}}
=
-\bm{\nabla_{r}}^T M \bm{\nabla}_{r} 
+ 
\sum_{i=1}^{\npart} \sum_{j>i}^{\npart} \frac{q_iq_j}{\left|\bm{r}_i-\bm{r}_j\right|}.
\end{equation}
where $\bm{\nabla}_{r}=\left(\bm{\nabla}_{r_1},\ldots,\bm{\nabla}_{\boldsymbol{r}_{\npart}}\right)^T$ 
collects the 3-dimensional Nabla operators for each particle 
and the diagonal $M_{ij}=\delta_{ij}\frac{1}{2\,m_i}$ matrix, which 
absorbs the $\frac{1}{2}$ term to shorten later notation.

Then, we consider a linear transformation of the coordinates:
\begin{equation}
\label{prima-transf}
U_x\bm{r}=\left(\bm{x}_1,\bm{x}_2,\ldots,\bm{x}_{\npart-1},\bm{x}_{\text{CM}}\right)^T
\end{equation}
in which the $\bm{x}_{\text{CM}}=\sum_{i=1}^{\npart}m_i\boldsymbol{r}_i/(\sum_{i=1}^{\npart}m_i)$ 
center-of-mass Cartesian coordinates (CMCC) are introduced and $(\bm{x}_1,\ldots,\bm{x}_{\npart-1})$
labels the translationally invariant Cartesian coordinates (TICC) corresponding to $U_x$.
Any transformation matrix $U_x$ can be selected which satisfy the translational invariance 
and the center-of-mass translational conditions:
\begin{align}
\label{Uconstraint1}
\sum_{j=1}^{\npart}(U_x)_{ij}=0 \hspace{0.5cm} 
{\text{with}} \hspace{0.5cm} i\in\left\lbrace 1,\ldots,\npart-1\right\rbrace, 
\end{align}
and 
\begin{align}
\label{Uconstraint2}
(U_x)_{N_p,j}=\frac{m_j}{m_{1\dots N_p}},
\end{align}
respectively, and $m_{j\ldots k}=\sum_{i=j}^km_i$. 

\begin{figure}[h]
  \centering
  \includegraphics[width=0.80\textwidth]{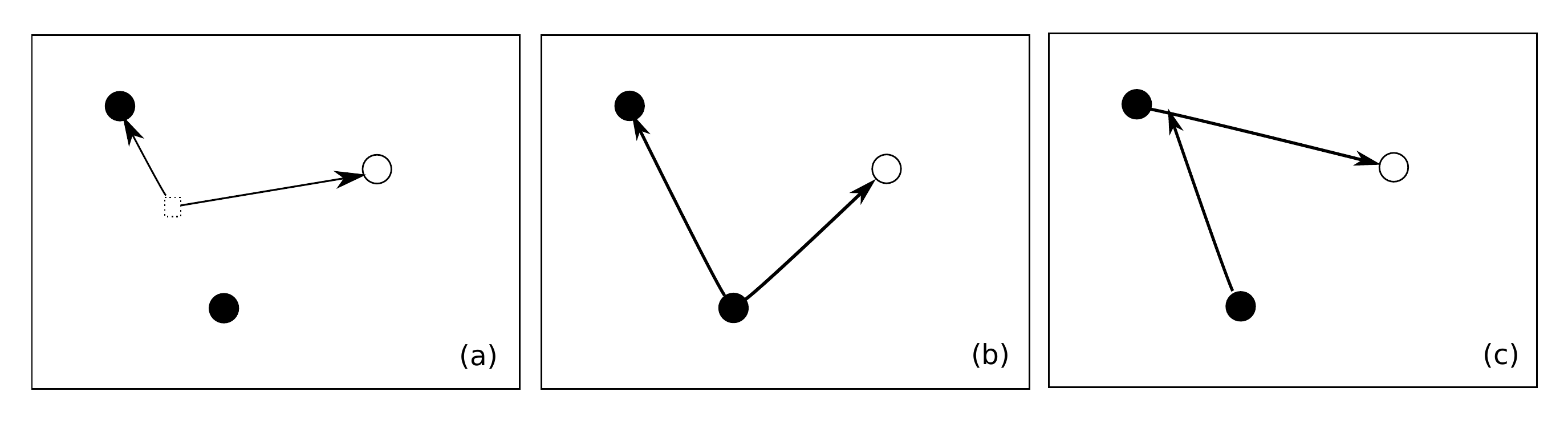}%
  \hfill
  \caption{\label{FIG:TICCs}
           Examples for translationally invariant Cartesian coordinates for a three particle system. 
           (a) center-of-mass-centered coordinates (CMC): the dashed box represent the center of mass; 
           (b) heavy-particle-centered coordinates (HPC); and 
           (c) a particular set of Jacobian coordinates. }
\end{figure}

There are infinitely many possible linear transformations which satisfy 
Eqs.~(\ref{Uconstraint1})--(\ref{Uconstraint2})
among which there are a few more common ones (Fig.~\ref{FIG:TICCs} visualizes three examples).
In the present work, we shall use
Jacobi coordinates,
\begin{align}
\label{eq:Jac-TICC}
\bm{x}_{i}^{\text{Jac}}=\sum_{j=1}^{i}\frac{m_{j}}{m_{1\ldots i}}\bm{r}_{j}-\bm{r}_{i+1},
\end{align}
the heavy-particle centered (HPC) coordinates (where the ``heavy particle'', 
$\bm{r}_\text{HP}$, 
is arbitrarily selected from the heavy particles)
\begin{align}
\label{eq:HPC-TICC}
\bm{x}_{i}^{\text{HPC}}=\bm{r}_{i}-\bm{r}_{\text{HP}},
\end{align}
and the center-of-mass-centered (CMC) coordinates
\begin{align}
\label{eq:CMC-TICC}
\bm{x}_{i}^{\text{CMC}}=\bm{r}_{i}-\sum_{j=1}^{\npart}\frac{m_{j}}{m_{i\ldots \npart}}\bm{r}_{j}.
\end{align}

Upon the transformation $\boldsymbol{r}\rightarrow(\boldsymbol{x}_1,\boldsymbol{x}_2,\ldots,\boldsymbol{x}_{\text{CM}})$ 
the operators change as follows: 
\begin{align}
\bm{\nabla}_{r} 
\rightarrow
\left(\bm{\nabla}_{x_1},\ldots,\bm{\nabla}_{x_{N_p-1}},\bm{\nabla}_{x_{\text{CM}}}\right)
\end{align}
where $(\bm{\nabla}_{\boldsymbol{x}_i})=\frac{\partial}{\partial x_{ia}}\ (a=x,y,z)$ 
and
\begin{equation}
\label{nonrel-H-transf}
\hat{T}=-\frac{1}{2\,m_{1\ldots \npart}}\nabla_{\bm{x}_{\text{CM}}}^2 - \nabla_{\bm{x}}^\text{T} \, \mu \, \nabla_{\bm{x}}
\end{equation}
Accordingly the mass matrix changes to (see also Sec.~\ref{SEC:constraints}),
\begin{equation}
U_x^{-\text{T}}MU_x^{-1}
=
\left[%
\begin{array}{cc}
\mu & 0 \\
0 & \frac{1}{2\,m_{1\ldots \npart}}
\end{array}
\right].
\end{equation}

\section{Explicitly correlated Gaussian functions}
\noindent 
Let us consider the 
family of
square-integrable, positive definite 
functions
\begin{equation}
\phi_I\left(\bm{r};\{\omega_I\}\right) : \mathbb{R}^{3N_p}\longrightarrow\mathbb{R}
\end{equation}
with $\left\{\omega_I\right\}$ parameters 
and $\text{dim}\left\{\omega_I\right\}\ge 1$.
We always choose
$\omega_{I,1}=A^{(q)}$, with 
a real ($3N_p\times 3N_p$) $A^{(q)}$ matrix of scalar values defined as
\begin{align}
\label{validDef1}
A^{(q)}=\bar{A}^{(q)}\otimes\mathbb{I}_3.
\end{align}
The superscript $q\in\{r,x,y,\ldots\}$ labels the coordinate set:
$q=r$ indicates that the matrix is expressed in LFCC,
otherwise $q=x,y,\ldots$ refers to a certain TICC selection.
The function $\phi_I$ keeps its mathematical form during the course of
the coordinate transformation and is parameterized with some $\left\{\omega_I\right\}$ set.
Upon a linear transformation $\bm{r}\rightarrow\bm{x}$, 
described by the matrix $U_x$, Eqs.~(\ref{prima-transf})--(\ref{Uconstraint2}),
the parameter set $\left\{\omega_I\right\}$ is also transformed as
\begin{align}
\label{validDef2}
\bar{A}^{(x)}=U_x^{-T}\bar{A}^{(r)}U_x^{-1}
\end{align}
with
\begin{align}
\label{validDef3}
\bar{A}^{(x)}=\left[ \begin{array}{cc} \mathcal{A}^{(x)} & 0 \\ 0 & c_A \end{array} \right]
\end{align}
where the $\mathcal{A}^{(x)}$ $\in\mathbb{R}^{(N_p-1)\times (N_p-1)}$ matrix corresponds 
to the selected TICC and 
$c_A$ is the only parameter related to the center-of-mass coordinates.

Using this family of functions, 
we approximate the exact eigenfunction of the Schr\"odinger equation with a linear combination
of $N_b$ properly (anti)symmetrized products of $\phi_I$ spatial and 
$\bm{\chi}_I^{S,M_S}$ spin functions: 
\begin{equation}
\label{wavefun}
\Psi(\mathbf{r})=\sum_{I=1}^{N_b}c_{I} \, \bm{\chi}_{I}^{S,M_{S}} \, \hat{Y}\phi_{I}\big(\mathbf{r};\{\omega_I\}\big)
\end{equation}
where the $c_I$'s are the linear combination coefficients and 
$\hat{Y}$ is the Young operator projecting onto the appropriate (anti)symmetric subspace.

In this work, we shall consider three types of ECG functions for the $\phi_I$ spatial basis function.  
These functions are introduced in the following subsections.

\subsection{Plain Explicitly correlated Gaussian functions (\texttt{pECG}s)} \label{subsec:plainECG}

The plain ECG functions (pECGs) are the simplest representatives of ECG-type functions:
\begin{equation}
\label{ecg}
\phi_I^{\texttt{pECG}}\left(\bm{r};A_I^{(r)}\right)=\exp\left[-\frac{1}{2}\bm{r}^TA_I^{(r)}\bm{r}\right].
\end{equation}
They are eigenfunctions of the square of the total angular momentum operator, $\hat{N}^2$, with $N=0$ quantum number
and they are parity eigenstates with $p=+1$.
The pECGs have simple analytic integral expressions for the most important operators.

\subsection{Floating explicitly correlated Gaussian functions (\texttt{FECG}s)} \label{subsec:FECG}
A more flexible functional form 
is introduced by allowing shifted particle positions ($\bm{r}-\bm{s}_I$)---%
hence the name floating ECG (FECG) functions---defined as:
\begin{align}
\phi_I^{\texttt{FECG}} \left(\bm{r};A_I^{(r)},\bm{s}_I^{(r)}\right) 
&= \exp\left[-\left(\bm{r}-\bm{s}_I^{(r)}\right)^TA_I^{(r)}\left(\bm{r}-\bm{s}_I^{(r)}\right)\right] \nonumber \\
&= \exp\left[-{\bm{s}_I^{(r)}}^TA_I^{(r)}\bm{s}_I^{(r)}-\bm{r}^TA_I^{(r)}\bm{r}+2\bm{r}^TA_I^{(r)}\bm{s}_I^{(r)}\right].
\end{align}
For non-vanishing $\bm{s}_I^{(r)}$ shift vectors, 
the \texttt{FECG}s are generally neither eigenfunctions of the total angular momentum operators, 
$\hat{N}^2$ and $\hat{N}_z$, nor eigenfunctions of 
the space-inversion operator.
Therefore, 
\texttt{FECG}s are usually considered to be less appropriate 
for approximating spherically symmetric states
than \texttt{pECG}s. At the same time, 
they are better suited for describing less delocalized particles (e.g., atomic nuclei)
due to the more flexible parameterization.
In a variational computation, the spherical symmetry is restored numerically by 
variationally optimizing basis sets of increasing size.

\subsection{Explicitly correlated Gaussian functions with global vector representation (\texttt{ECGs-GVR})} \label{subsec:ECGGVR}
The ideal basis functions are eigenfunctions of the spatial symmetry operators 
($\hat{\bm{N}}^2$, $\hat{N}_z$ and parity)
and they are sufficiently flexible in their parameterization 
to account for very different types of particle distributions.

The \texttt{pECG} functions can be made eigenfunctions of $\hat{N}^2$ by multiplying it with 
an angular factor $\theta_{NM_N}(\bm{r})$ defined as a vector-coupled product of solid spherical 
harmonics $\mathcal{Y}_{l}(\bm{r}_i)$ of particle $i$
\begin{equation}
\tilde\theta_{NM_N}(\bm{r}) = 
\left[\left[\left[\mathcal{Y}_{l_1}(\bm{r}_1)\mathcal{Y}_{l_2}(\bm{r}_2)\right]_{N_{12}}\mathcal{Y}_{l_3}(\bm{x}_3)\right]_{N_{123}}\ldots\right]_{N\,M_N}.
\label{eq:veccoup}
\end{equation}
The subsystems' angular momenta, $l_1,l_2,\ldots,N_{12},N_{123},\ldots$ are not conserved and 
for a realistic description of few-body problems one must 
include several ($l_1,l_2,\ldots ,l_{N};N_{12},N_{123},\ldots$) sets.
The various possible partial wave contributions from each set increase both the basis set dimension and 
the complexity of the integrals associated
with expectation values of quantum mechanical operators. 
Moreover, the change of $\theta_{NM_N}(\bm{r})$ 
upon changing the coordinate set usually leads to very complicated expressions.

An alternative to this commonly used vector-coupled form 
has been proposed 
by Suzuki and Varga \cite{suzukivarga1998,suzukivarga2008}:
\begin{equation}
\label{globalvector}
\theta_{NM_N}\left(\bm{r};\bm{u}^{(r)},K\right)=\left|\bm{v}^{(r)}\right|^{2K+L}Y_{NM_N}
\end{equation}
with the {\it global vector} $\bm{v}\equiv\sum_{i=1}^Nu_i\bm{r}_i=\tilde{\bm{u}}^{(r)}\bm{r}$ being a linear 
combination of all (pseudo-)particle coordinates. It has been shown that when used in a variational
procedure the pECGs multiplied with either Eq.~(\ref{eq:veccoup}) or (\ref{globalvector}) 
result in a mathematically equivalent representation.
In Eq.~(\ref{globalvector}) only the conserved total orbital angular momentum quantum number, $N$, appears, whereas
the explicit coupling of the subsystems' angular momenta is completely avoided
(it is implicitly carried by the global vectors in the variational ansatz).
The coefficients $u_i$ in the global vector are variational parameters to be optimized by minimizing the energy.
Upon the transformation of the coordinates, Eq.~(\ref{prima-transf})--(\ref{Uconstraint2}),
the vector $\bm{u}\in\mathbb{R}^{\npart}$ transforms as
\begin{equation}
\label{validDef4}
U_x^{-T}\bm{u}^{(r)} = \left( \begin{array}{c} \bm{u}' \\ c_U \end{array} \right),
\end{equation}
The $K$ integer parameter in Eq.~(\ref{globalvector}) introduces additional variational flexibility 
for the basis function (qualitatively, it helps to describe more efficiently localized, 
vibrating atomic nuclei).

In the so-called global vector representation (GVR)
the angular term in Eq.~(\ref{globalvector}) multiplies a \texttt{pECG}:
\begin{equation}
\label{ecg-gvr}
\phi^{\texttt{ECG-GVR}\,[NM_N]}_I\left(\bm{r};A_I^{(r)},\bm{u}_I^{(r)},K_I\right)=
\theta_{NM_N}\left(\bm{r};\bm{u}^{(r)},K_I\right) \cdot\exp\left[-\frac{1}{2}\bm{r}^TA_I^{(r)}\bm{r}\right].
\end{equation}
This choice of the basis functions 
leads to compact $\npart$-particle analytic integrals for the most important physical operators 
and corresponds to well-defined 
values for the spatial quantum numbers (total angular momentum, $N$ and $M_N$, and parity).

\section{Identification of the global translational contributions in an LFCC calculation}\label{SEC:constraints}

In this section we study the analytic integrals of the overlap and the Schr\"odinger Hamiltonian
expressed in the original laboratory-frame Cartesian coordinates (LFCC)
in order to identify the translationally-invariant terms and eliminate others, which originate from
the overall translation of the system.

Most importantly, we rely on the invariance of the functional form of all ECG-type functions considered in this work, 
upon a linear transformation, and in particular the Eq.~(\ref{prima-transf})--(\ref{Uconstraint2}) transformation,
of the coordinates. The mathematical form of the functions is unchanged, while the parameters transform as follows:
\begin{align}
\label{ecg-transf}
\phi^{\texttt{ECG}}_I\left(\bm{r}\rightarrow U_x^{-1}\bm{x};A_I^{(r)}\right) 
&= \exp\left[-\frac{1}{2}\left(U_x^{-1}\bm{x}\right)^TA_I^{(r)}\left(U_x^{-1}\bm{x}\right)\right] \nonumber \\
&= \exp\left[-\frac{1}{2}\bm{x}^T\left(U_x^{-T}\bar{A}_I^{(r)}U_x^{-1}\otimes I_3\right)\bm{x}\right] \nonumber \\
&= \exp\left[-\frac{1}{2}\bm{x}^TA_I^{(x)}\bm{x}\right]=\phi^{\texttt{ECG}}\left(\bm{x};A_I^{(x)}\right).
\end{align}

Conceptually, a special parameterization of the basis functions allows for 
the detection and removal of CM translational contributions 
at the level of the kinetic energy expectation value.
In our earlier work \cite{Benjamin2013}, we have pointed out that a few controllable CM-dependent 
terms and factors can be identified in the (kinetic energy) integral expressions. 
These terms were eliminated during the course of the integral evaluation
in order to obtain translation-free values.

Ref.~\cite{Benjamin2013} focused on \texttt{ECG-GVR} functions in which 
the variational parameter matrix $A_I$ and the global vector $\bm{u}_I$ 
was transformed back and forth between different coordinate representations according to
\begin{equation}
\label{eq:backandforthtransfonA}
\bar{A}_I^{(r)}=U_{x}^T\bar{A}_I^{(x)}U_{x} \hspace{0.5cm}\Longleftrightarrow\hspace{0.4cm} \bar{A}_I^{(x)}=U_{x}^{-T}\bar{A}_I^{(r)}U_{x}^{-1},
\end{equation}
and
\begin{equation}
\label{eq:backandforthtransfonu}
\bm{u}_I^{(r)}=U_x^T\bm{u}_I^{(x)} \hspace{0.5cm}\Longleftrightarrow\hspace{0.4cm} \bm{u}_I^{(x)}=U_x^{-T}\bm{u}_I^{(r)}
\end{equation}
where $U_x$ satisfies the translational invariance and CM conditions, 
Eqs.~(\ref{Uconstraint1}) and (\ref{Uconstraint2}), respectively.

When expressed with some TICC (and CMCC) 
$A_I$ and $\bm{u}_I$ have the special block structure: 
\begin{equation}
A_I^{(x)} = \left( \begin{array}{cc} \mathcal{A}_I^{(x)} & 0 \\ 0 & c_A \end{array} \right) \hspace{0.5cm}{\text{and}}\hspace{0.4cm}  
\bm{u}_I^{(x)} = \left( \begin{array}{c} {\bm{u}}'_I \\ c_U \end{array} \right) .
\end{equation}
Since $c_A$ and $c_U$ are related to the CM coordinates,
$\bm{x}_{\text{CM}}$, the system is ``at rest'' only for $c_A=0$ and $c_U=0$.
Although $c_U$ can be set to zero without any problems, 
if $c_A$ was chosen to be zero, the $A$ matrix would become singular, which violates
the square integrable and positive definiteness requirements for the basis functions.

Ref.~\cite{Benjamin2013} defined the following approach
to handle the $c_A$-dependent terms without violating the square-integrability and positive-definiteness
conditions: 
\begin{enumerate}
\item For each basis function $I$, generate, optimize, 
      or read in the $(\bar{A}^{(x)}_{I})_{ij}$ values 
      with $i,j\in\left\lbrace 1,\ldots,N_\text{b}\right\rbrace$. 
\item Construct the elements of the exponent matrix in the LFCC framework as
      \begin{equation}
      \label{benjparametr}
      (\bar{A}^{(r)}_I)_{ij}=-(\bar{A}^{(x)}_I)_{ij}\left(1-\delta_{ij}\right)+\left(\sum_{k=1,k\ne i}^{N_p}(\bar{A}^{(x)}_I)_{ij}\right)\delta_{ij}
      +c_A\frac{m_i}{m_{1\ldots N_p}}\frac{m_j}{m_{1\ldots N_p}}
      \end{equation}
      with $i,j=1,\ldots,N_p$ and some $c_A>0$ value.
\item For $c_A>0$ the matrices $\bar{A}_I^{(r)}$ are non-singular, 
      $|\bar{A}_I^{(r)}|$ and also $\bar{A}_I^{(r)^{-1}}$ can be evaluated.
      At the same time, the total kinetic energy contains some translational 
      effects (``contamination'').
\item It was shown in Ref.~\cite{Benjamin2013} that the only CM-dependent term arising in the analytic kinetic energy integral is 
      the $R_{IJ}$ term defined in Eq.~(32) of Ref.~\cite{Benjamin2013}:
      \begin{align}
      R_{IJ} =& \frac{3}{2} \Tr\left[A_{IJ}^{(r)^{-1}}A_J^{(r)}MA_I^{(r)}\right] \nonumber \\
      =& \frac{3}{2} \Tr\left[\left(A_{IJ}^{(x)}\right)^{-1}A_J^{(x)}U_xMU^T_xA_I^{(x)}\right] \nonumber \\
      =& \frac{3}{2} \Tr\left[\left(\mathcal{A}_{IJ}^{(x)}\right)^{-1}\mathcal{A}_J^{(x)}\mu^{(x)}\mathcal{A}_I^{(x)}\right]+\frac{3}{4}c_Ac_M \nonumber
      \end{align}
      Then, the translational contamination was eliminated by 
      replacing $R_{IJ}$, with $R_{IJ}-3c_A/\left(4m_{1\ldots N_p}\right)$ in the
      expression of the kinetic energy matrix element (see 
      Eqs.~(33)--(37) of Ref.~\cite{Benjamin2013}).   
\end{enumerate}

At this point, we mention
that the parameterization of the $A^{(r)}$ matrix expressed in Eq.~(\ref{benjparametr}) is the algebraic computation of 
the backward transformation from a specific TICC, namely the CMC coordinate set introduced in Eq.~(\ref{eq:CMC-TICC})
to LFCC.
This scheme therefore forces the $A^{(r)}$ matrix to be obtained from the block diagonal $A^{(x)}$ form through 
a specific mapping (a specific $U_x$ transformation matrix).

As to the generalization of this approach, we note that 
one can build more general schemes in which $A^{(x)}$ is mapped to the $A^{(r)}$ matrix by various 
transformations $U_{a}$, $a\in\left[x,y,,z,\ldots\right]$ in order to enhance the flexibility of 
the basis functions, and thereby to gain direct access to a broader region in the physical parameter space. 
The idea is related to the multi-coordinate or {\it multi-channel} optimization of Suzuki and Varga \cite{suzukivarga},
also discussed by M\'atyus \cite{Matyus2012}.

The present work generalizes the elimination approach of \cite{Benjamin2013},
summarized in Steps 1.--4.,
for the case of the multi-coordinate / multi-channel optimization.
For this purpose we work out a fundamental relationships of the integral expressions
corresponding to basis functions parameterized in different coordinate sets
(defined by different $U_{x}$ and $U_{y}$ transformation matrices).

% \subsection{Required relations}
\subsection{Fundamental relationships\label{ch:fundam}}

First, we establish two mathematical relations that will 
be crucial in the extraction of $c_A$-dependent terms:
\begin{align}
\label{1genconstr}
U_{x}\bar{A}_{IJ}^{-1}U_{y}^T &= \left[
\begin{array}{cc}
\mathcal{A}_{IJ}^{-1} & 0 \\
0 & \frac{1}{2 c_{A}}
\end{array}
\right]
\end{align}
and
\begin{align}
\label{2genconstr}
U_{y}MU_{x}^T &= \left[
\begin{array}{cc}
\mu & 0 \\
0 & \frac{c_M}{2}
\end{array}
\right],
\end{align}
with $U_{x}$ and $U_{y}$ being the transformation matrices associated 
with two different TICC sets,
Eqs.~(\ref{prima-transf})--(\ref{Uconstraint2}), for 
a pair of function $\phi_I$ and $\phi_J$, respectively. 
$\mathcal{A}_{IJ}$ and $\mu$ are square matrices of dimension $N_p-1$.
$c_{A}$ is a free parameter and $c_M\equiv \frac{1}{2m_{1\ldots N_p}}$ as will be determined below.
$\bar{A}_{IJ}$ is an ($N_p\times N_p$) matrix obtained as a sum of
the $\bar{A}^{(r)}$ matrices of $\phi_I$ and $\phi_J$:
\begin{equation}
\bar{A}_{IJ} = \bar{A}^{(r)}_I + \bar{A}^{(r)}_J = U_{x}^{T}\bar{A}^{(x)}_IU_{x}+U_{y}^{T}\bar{A}^{(y)}_JU_{y}.
\end{equation}
For later convenience, we write Eq.~(\ref{1genconstr}) in a different form:
\begin{align}
U_{x}\bar{A}_{IJ}^{-1}U_{y}^T =& \left(U_{y}^{-T}\bar{A}_{IJ}U_{x}^{-1}\right)^{-1} 
  = \left[
  \begin{array}{cc}
  \mathcal{A}_{IJ} & 0 \\
  0 & 2 c_{A}
  \end{array}
  \right]^{-1}
\end{align}
and
\begin{align}
U_{y}^{-T}\bar{A}_{IJ}U_{x}^{-1} =& U_{y}^{-T}\left(U_{x}^T\bar{A}^{(x)}_IU_{x}+U_{y}^T\bar{A}^{(y)}_JU_{y}\right)U_{x}^{-1} \nonumber \\
=& U_{y}^{-T}U_{x}^T\bar{A}^{(x)}_I+\bar{A}^{(y)}_JU_{y}U_{x}^{-1}.
\end{align}
So, we need to prove 
\begin{align}
U_yU_x^{-1} &= \left[
  \begin{array}{cc}
  \mathcal{U} & 0 \\
  0 & 1
  \end{array}
  \right],
\label{1recasted}
\end{align}
to show the validity of Eq.~(\ref{1genconstr}).
In simple terms, Eq.~(\ref{1recasted}) means that 
the space of TICCs is closed: any linear combination of TICC coordinates is
also a TICC coordinate (no contamination from the CMCC). 
It is easy to see qualitatively that this statement
should be correct. The following equations provide the mathematical proof of it.
During the derivation, we shall rely only on the 
properties of a $U$ matrix, Eqs.~(\ref{Uconstraint1}) and (\ref{Uconstraint2}), 
and general mathematical
properties of determinants.

Let us consider $\left(U_y\right)_{ab}\left(U_x^{-1}\right)_{bc}$ with 
$U^{-1}=\frac{1}{\det(U)}\adj(U)$ and 
$\adj(U)=C^T$ is the transpose of the cofactor matrix.
Then,
\begin{equation}
\label{1middle}
\left(U^{-1}\right)_{iN_p}=\frac{C_{N_pi}}{U_{N_p1}C_{N_p1}+\ldots+U_{N_pN_p}C_{N_pN_p}}.
\end{equation}
Due to Eq.~(\ref{Uconstraint1}), 
\begin{equation}
C_{N_pi}=\det\left(
\begin{array}{cccccc}
U_{11}      & \cdots & U_{1\, i-1}   & U_{1\, i+1} & \cdots & \left(-U_{11}-U_{12}+\ldots -U_{1N_p}\right)                     \\
\vdots      &        & \vdots        & \vdots      &        & \vdots                                                           \\
U_{N_p-1\, 1} & \cdots & U_{N_p-1\, i-1} & U_{1\, i+1} & \cdots & \left(-U_{N_p-1\, 1}-U_{N_p-1\, 2}+\ldots -U_{N_p-1\, N_p}\right)
\end{array}
\right),
\end{equation}
and hence $C_{N_p1}=C_{N_p2}=\ldots=C_{N_pN_p}$. Moreover, we also have from Eq.~(\ref{Uconstraint2}) that
\begin{equation}
\label{eq:genconstr0}
\left(U^{-1}\right)_{iN_p}=\frac{C_{N_pi}}{\left(U_{N_p1}+\ldots+U_{N_pN_p}\right)C_{N_pi}}=1,
\end{equation}
from which we obtain $\det(U)=C_{N_pi}$.
From Eqs.~(\ref{Uconstraint2}) and (\ref{eq:genconstr0}), it follows that
\begin{align}
\sum_{b=0}^{N_p} \left(U_y\right)_{N_pb}\left(U_x^{-1}\right)_{bN_p}=1
\end{align}
and
\begin{align}
\sum_{b=0}^{N_p} \left(U_y\right)_{ib}\left(U_x^{-1}\right)_{bN_p}=0 & \hspace{0.5cm} 
\text{for} \hspace{0.2cm} i\in\left\lbrace0,N_p-1\right\rbrace.
\end{align}
To complete the proof, we need to show that
\begin{equation}
\label{1adv}
\sum_{a=0}^{N_p} \left(U_y\right)_{N_pa}\left(U^{-1}\right)_{ai} = 0 \hspace{0.5cm} 
\text{for} \hspace{0.2cm} i\in\left\lbrace0,\ldots,N_p-1\right\rbrace,
\end{equation}
which is rewritten using Eq.~(\ref{Uconstraint2}) as:
\begin{equation}
\frac{m_1}{m_{1\ldots N_p}}\frac{C_{21}}{C_{N_pi}}+\frac{m_2}{m_{1\ldots N_p}}\frac{C_{22}}{C_{N_pi}}+\ldots = 
\frac{1}{C_{N_pi}} \left(\frac{m_1}{m_{1\ldots N_p}}C_{21}+\frac{m_2}{m_{1\ldots N_p}}C_{22}+\ldots\right)\overset{!}{=}0,
\end{equation}
where the term in the parenthesis is zero, because it is the determinant of a matrix with two identical rows.
With this result, we have verified Eq.~(\ref{1genconstr}).

\vspace{0.4cm}

Next, we give the proof of Eq.~(\ref{2genconstr}) by investigating $U_yMU_x^T$ element by element:
\begin{align}
\left(U_yMU_x^T\right)_{ij} &= \sum_{k,l}\left(U_y\right)_{ik}M_{kl}\left(U_x^T\right)_{lj} \nonumber \\
                            &= \sum_k \left(U_y\right)_{ik}\left(U_x^T\right)_{kj}\frac{1}{2m_k} \nonumber \\
                            &= \sum_k \left(U_y\right)_{ik}\left(U_x\right)_{jk}\frac{1}{2m_k}.
\end{align}

We can separate three cases 
\begin{align}
\label{endof2}
& \sum_{k} \frac{m_k}{m_{1\ldots N_p}}\frac{m_k}{m_{1\ldots N_p}}\frac{1}{2m_k} = \sum_k \frac{m_k}{2m_{1\ldots N_p}^2} = 
\frac{1}{2m_{1\ldots N_p}} = \frac{c_M}{2} \hspace{0.4cm} 
{\text{for}} \hspace{0.4cm} i=N_p \wedge j=N_p \hspace{0.1cm}, \nonumber \\
& \sum_{k} \frac{m_k}{m_{1\ldots N_p}}\frac{1}{2m_k} (U_x)_{jk}=0 \hspace{0.4cm} 
{\text{for}} \hspace{0.4cm} i=N_p \wedge j\in \left\lbrace 1,\ldots,N_p-1\right\rbrace \hspace{0.1cm}, \\
& \sum_{k} (U_y)_{ik}\frac{m_k}{m_{1\ldots N_p}}\frac{1}{2m_k}=0 \hspace{0.4cm} 
{\text{for}} \hspace{0.4cm} i\in \left\lbrace 1,\ldots,N_p-1\right\rbrace \wedge j=N_p \hspace{0.1cm}, \nonumber
\end{align}
which completes the proof of Eq.~(\ref{2genconstr}).

Using the two fundamental relations, Eqs.~(\ref{1genconstr}) and (\ref{2genconstr}), 
which we have just verified, we proceed to the identification
of the CM-related terms in the integral expressions for
the three types of ECG functions introduced in Section~\ref{subsec:ECGGVR}.

%\subsection{Generalized constraint: \texttt{pECG}}
\subsection{Translationally invariant expressions for the \texttt{pECG}-type functions}
The matrix element of the kinetic energy operator for \texttt{pECG}-type functions is
\begin{equation}
T_{IJ}=\frac{\left\langle\phi_I\left|\bm{\nabla}_{\bm{r}}^TM\bm{\nabla}_{\bm{r}}\right|\phi_J\right\rangle}{\left|\phi_I\right|\,\left|\phi_J\right|} =
\frac{\left\langle\phi_I|\phi_J\right\rangle}{\left(\left\langle\phi_I|\phi_I\right\rangle\left\langle\phi_J|\phi_J\right\rangle\right)^{\frac{1}{2}}}
\cdot 6 \, \underset{\equiv R}{\underbrace{\Tr\left(\bar{A}_{IJ}^{-1}\bar{A}_I^{(r)}M\bar{A}_J^{(r)}\right)}}
\end{equation}
The $R$ term encompasses the total kinetic energy corresponding to the $IJ$-th matrix element and account for all particles. 
We investigate this term and isolate $c_A$ contributions in order to eliminate 
the center-of-mass kinetic energy contributions.
The $c_A$-dependent terms cancel in the overlap integrals (see Appendix \ref{APP:cA_det_terms}),
so, using Eqs.~(\ref{1genconstr}) and (\ref{2genconstr}),  we write
\begin{align}
\label{ecggencon}
R = & \Tr\left(A_{IJ}^{-1}U_x^TA_I^{(x)}U_xMU_y^TA_J^{(y)}U_y\right) \nonumber \\
= & \Tr\left[
\left( \begin{array}{cc} \mathcal{A}_{IJ}^{-1} & 0 \\ 0 & \frac{1}{2c_{A}} \end{array} \right)
\left( \begin{array}{cc} \mathcal{A}_I^{(x)} & 0 \\ 0 & c_{A} \end{array} \right)
\left( \begin{array}{cc} \mu & 0 \\ 0 & \frac{c_M}{2} \end{array} \right)
\left( \begin{array}{cc} \mathcal{A}_J^{(x)} & 0 \\ 0 & c_{A} \end{array} \right)
\right) ,
\end{align}
Thereby, the contributions related to the overall translation are eliminated by subtracting the
$c_{A}$ and $c_M$ dependent term(s):
\begin{equation}
\label{ecg-transl}
R^{(\text{TI})} 
= R - \frac{1}{2}c_Mc_A 
\end{equation}
where the superscript TI refers to ``translationally invariant'' and $c_M$ has been 
introduced in Eq.~(\ref{endof2}).

% \subsection{Generalized constraint: \texttt{FECG}}
\subsection{Translationally invariant expressions for the \texttt{FECG}-type functions}

The kinetic energy matrix element for \texttt{FECG}-type basis functions is \cite{FECG-integrals-advances1978}:
\begin{align}
T_{IJ} = & \frac{\left\langle\phi_I|\phi_J\right\rangle}{\left(\left\langle\phi_I|\phi_I\right\rangle\left\langle\phi_J|\phi_J\right\rangle\right)^{\frac{1}{2}}} 
\cdot \Big[ \underset{\equiv Q}{\underbrace{4\left(\bm{s}-\bm{s}_I\right)^TA_I^{(r)}MA_J^{(r)}\left(\bm{s}-\bm{s}_J\right)}} 
+ 6\cdot \underset{\equiv R}{\underbrace{\Tr\left(M\bar{A}_J^{(r)}\bar{A}_{IJ}^{-1}\bar{A}_I^{(r)}\right)}} \big]
\end{align}
where $\bm{s}=A_{IJ}^{-1}\left(A_I^{(r)}\bm{s}_I+A_J^{(r)}\bm{s}_J\right)$ and every $\bm{s}$ vector is expressed in 
the LFCC set (the superscripts $^{(r)}$ have been omitted for clarity).
For the $R$ term, we use the result obtained from the \texttt{pECG} functions, Eq.~(\ref{ecg-transl}), 
so we need to consider the $Q$ term.
First of all, we notice that:
\begin{equation}
4\left(\bm{s}-\bm{s}_I\right)^TA_I^{(r)}MA_J^{(r)}\left(\bm{s}-\bm{s}_J\right)=
4\left(\bm{s}_I-\bm{s}_J\right)^TA_J^{(r)}A_{IJ}^{-1}A_I^{(r)}MA_J^{(r)}A_{IJ}^{-1}A_I^{(r)}\left(\bm{s}_J-\bm{s}_I\right),
\end{equation}
and thereby
\begin{align}
Q = & 4\left(\bm{s}_I-\bm{s}_J\right)^TA^{(r)}_JA_{IJ}^{-1}A^{(r)}_IMA^{(r)}_JA_{IJ}^{-1}A^{(r)}_I\left(\bm{s}_J-\bm{s}_I\right) \nonumber \\
= & 4\left(\bm{s}_I-\bm{s}_J\right)^T \, \left[\left( U_x^T\bar{A}_I^{(x)}U_x \,\bar{A}_{IJ}^{-1} \,U_y^T\bar{A}_J^{(y)}U_y \,M \,U_x^T\bar{A}_I^{(x)}U_x \,\bar{A}_{IJ}^{-1} \,U_y^T\bar{A}_J^{(y)}U_y \right) \otimes \mathbb{I}_3 \right] \, \left(\bm{s}_J-\bm{s}_I\right) \nonumber \\
= & 4\left(\bm{s}_I-\bm{s}_J\right)^T \left[\left( U_x^T
\left( \begin{array}{cc} \mathcal{A}_I^{(x)} & 0 \\ 0 & c_{A} \end{array} \right)
\left( \begin{array}{cc} \mathcal{A}_{IJ}^{-1} & 0 \\ 0 & \frac{1}{2c_{A}} \end{array} \right)
\left( \begin{array}{cc} \mathcal{A}_J^{(y)} & 0 \\ 0 & c_{A} \end{array} \right)
\left( \begin{array}{cc} \mu & 0 \\ 0 & \frac{c_M}{2} \end{array} \right) \right. \right. \nonumber \\
& \left. \left. \left( \begin{array}{cc} \mathcal{A}_I^{(x)} & 0 \\ 0 & c_{A} \end{array} \right)
\left( \begin{array}{cc} \mathcal{A}_{IJ}^{-1} & 0 \\ 0 & \frac{1}{2c_{A}} \end{array} \right)
\left( \begin{array}{cc} \mathcal{A}_J^{(y)} & 0 \\ 0 & c_{A} \end{array} \right) U_y \right) \otimes \mathbb{I}_3 \right]
\left(\bm{s}_J-\bm{s}_I\right) ,
\end{align}
where Eqs.~(\ref{1genconstr}) and (\ref{2genconstr}) are used in the third step.
Finally, we identify the $c_A$-dependent terms in $Q$ as:
\begin{equation}
\label{GENCONSTR:fecg}
Q-Q^{\text{TI}}=
\frac{1}{2} \, c_{A} 
\, c_M \, \left(\bm{s}_I-\bm{s}_J\right) \left[ (U_x^T)_{iN} (U_y)_{Nj} \otimes \mathbb{I}_3 \right] \left(\bm{s}_J-\bm{s}_I\right)
\end{equation}
where $(U_q)_{Nj}$ was defined in Eq.~(\ref{Uconstraint2}).

% \subsection{Generalized constraint: \texttt{ECG-GVR}}
\subsection{Translationally invariant expressions for the \texttt{ECG-GVR}-type functions}
In this subsection, we consider the kinetic energy matrix element for ECG-GVR functions \cite{suzukivarga,Matyus2012}:
\begin{align}
T_{IJ} = & \left(\frac{\left|2A_I^{(r)}\right|^{\frac{1}{2}}\left|2A_J^{(r)}\right|^{\frac{1}{2}}}{\left|A_I^{(r)}+A_J^{(r)}\right|}\right)^{\frac{3}{2}} 
\left(\frac{p_{\bm{u}_I,\bm{u}_I}}{q_{\bm{u}_I}}\right)^{K_I} \left(\frac{p_{\bm{u}_J,\bm{u}_J}}{q_{\bm{u}_J}}\right)^{K_J} 
\left(\frac{p_{\bm{u}_I,\bm{u}_I}}{\sqrt{q_{\bm{u}_I}q_{\bm{u}_J}}}\right)^{L} \nonumber \\
& \times \sum_{m=0}^{\text{min}\left(K_I,K_J\right)} \left(\frac{p_{\bm{u}_I,\bm{u}_J}^2}{p_{\bm{u}_I,\bm{u}_I}p_{\bm{u}_J,\bm{u}_J}}\right)^{m} \times 
\left[\frac{3}{2}R+\left(K_I-m\right)\frac{P_{\bm{u}_I,\bm{u}_I}}{p_{\bm{u}_I,\bm{u}_I}}\right. \nonumber \\
& \left.+\left(K_J-m\right)\frac{P_{\bm{u}_J,\bm{u}_J}}{p_{\bm{u}_J,\bm{u}_J}}+\left(N+2m\right)\frac{P_{\bm{u}_I,\bm{u}_J}}{p_{\bm{u}_I,\bm{u}_J}}\right]
H_{NK_IK_Jm},
\label{eq:tkinecggvr}
\end{align}
where 
\begin{align}
\label{ecggvr-transl-1}
p_{\bm{u}_Q,\bm{u}_Z} &= \bm{u}_Q^T\bar{A}_{IJ}^{-1}\bm{u}_Z, \\
\label{ecggvr-transl-2}
P_{\bm{u}_I,\bm{u}_I} &= -\bm{u}_I^T\bar{A}_{IJ}^{-1}\bar{A}_J^{(r)}M\bar{A}_J^{(r)}\bar{A}_{IJ}^{-1}\bm{u}_I, \\
\label{ecggvr-transl-3}
P_{\bm{u}_J,\bm{u}_J} &= -\bm{u}_J^T\bar{A}_{IJ}^{-1}\bar{A}_I^{(r)}M\bar{A}_I^{(r)}\bar{A}_{IJ}^{-1}\bm{u}_J, \\
\label{ecggvr-transl-4}
P_{\bm{u}_I,\bm{u}_J} &=  \bm{u}_I^T\bar{A}_{IJ}^{-1}\bar{A}_J^{(r)}M\bar{A}_I^{(r)}\bar{A}_{IJ}^{-1}\bm{u}_J, \\
\label{ecggvr-transl-5}
q_{\bm{u}_Z} &= \frac{1}{2}\bm{u}_Z^T\bar{A}_Z^{(r)^{-1}}\bm{u}_Z, \\
\label{ecggvr-transl-6}
R &= \Tr\left[\bar{A}_{IJ}^{-1}\bar{A}_J^{(r)}M\bar{A}_I^{(r)}\right].
\end{align}
with $Q,Z\in\left\{I,J\right\}$ and $H_{NK_IK_Jm}$ is a set of precomputed values defined in Ref.~\cite{Matyus2012}.
Among these, only $R$ and $P$ terms arise from the application of the kinetic operator on the bra and the ket 
\texttt{ECG-GVR} functions.

In Ref.~\cite{Benjamin2013}, the constraint $c_U=0$ was introduced in order to facilitate the elimination
of CM contributions from the terms in Eqs.~(\ref{ecggvr-transl-1})-(\ref{ecggvr-transl-6}). 
Here we provide formulas for the elimination of CM kinetic energy that consider a non trivial value for $c_U$.

We calculate the corrections to the only terms generated by the kinetic energy integral 
on the generating functions (see Eq.~(S34) in Ref.~\cite{Matyus2012}), that is, 
$R$, $P_{\bm{u}_I,\bm{u}_I}$, $P_{\bm{u}_J,\bm{u}_J}$ and $P_{\bm{u}_I,\bm{u}_J}$.
Using Eqs.~(\ref{eq:backandforthtransfonA})--(\ref{eq:backandforthtransfonu}) and then
Eqs.~(\ref{1genconstr})--(\ref{2genconstr}) we write: 
\begin{align}
P_{\bm{u}_I,\bm{u}_J} =& \bm{u}_I^{(x)\,T}U_x\bar{A}_{IJ}^{-1}U_y^T\bar{A}_J^{(y)}U_yMU_x^T\bar{A}_I^{(x)}U_x\bar{A}_{IJ}^{-1}U_y^T\bm{u}_J^{(y)}  \nonumber \\
 =& \left( \begin{array}{cc} {\bm{u}}'_I & c_{U_I} \end{array} \right) 
    \left( \begin{array}{cc} \mathcal{A}_{IJ}^{-1} & 0 \\ 0 & \frac{1}{2c_{A}} \end{array} \right)
    \left( \begin{array}{cc} \mathcal{A}_J^{(y)} & 0 \\ 0 & c_{A} \end{array} \right)
    \left( \begin{array}{cc} \mu & 0 \\ 0 & c_M \end{array} \right) \nonumber \\
 &  \cdot \left( \begin{array}{cc} \mathcal{A}_I^{(x)} & 0 \\ 0 & c_{A} \end{array} \right)
          \left( \begin{array}{cc} \mathcal{A}_{IJ}^{-1} & 0 \\ 0 & \frac{1}{2c_{A}} \end{array} \right)
          \left( \begin{array}{c} {\bm{u}}'_J \\ c_{U_J} \end{array} \right)
\end{align}
and analogous expressions are obtained for $P_{\bm{u}_I,\bm{u}_I}$ and $P_{\bm{u}_I,\bm{u}_J}$.
As a result, the translationally invariant (TI) expressions are
\begin{align}
\label{GENCONSTR:ecggvr}
P_{\bm{u}_I,\bm{u}_J}^{\text{(TI)}} &= 
  P_{\bm{u}_I,\bm{u}_J} - \frac{1}{4} c_{U_I}c_Mc_{U_J} , \\ 
P_{\bm{u}_I,\bm{u}_I}^{\text{(TI)}} &= 
  P_{\bm{u}_I,\bm{u}_I} + \frac{1}{4} c_{U_I}c_Mc_{U_I} , \\
P_{\bm{u}_J,\bm{u}_J}^{\text{(TI)}} &= 
  P_{\bm{u}_J,\bm{u}_J} + \frac{1}{4} c_{U_J}c_Mc_{U_J} .
\end{align}
Furthermore, $R$ in Eq.~(\ref{eq:tkinecggvr}) is replaced with $R^{\text{TI}}$ given in Eq.~(\ref{ecg-transl}), 
which completes the list of expressions which will be used to eliminate the effect of
the overall translation in LFCC calculations carried out with the \texttt{ECG-GVR}-type functions.

\subsection{Multi-channel optimization}

The exact wave function is estimated as a linear combination of (anti)symmetrized products of spin and spatial
functions in a variational procedure. The linear combination coefficients are determined by solving the generalized eigenvalue
problem. In what follows, we shall discuss in detail how we parameterize the spatial basis functions. 
The spatial functions
are generated one after the other and (their parameters) are optimized variationally using the competitive selection procedure \cite{suzukivarga}.
In order to obtain very accurate energy levels, we repeatedly fine-tune the parameters of the selected basis functions
using Powell's method \cite{Powell04}. The convergence of the computed states is ensured by the variational principle.
As an additional check, we also calculate the virial ratio.

The efficiency of the optimization procedure can be enhanced by tuning the basis function parameters expressed in different
translationally invariant coordinate sets. 
Qualitatively speaking, different TICC sets describe efficiently different
``groupings'' of the particles (pairs and triples of particles, etc.). 
The basis functions which describe the interaction
of these pairs or groups of particles can be directly expressed in that particular TICC representation. 
So, the calculations (Hamiltonian representation, matrix elements, etc.) are performed in laboratory-fixed Cartesian coordinates,
but the optimization of the basis function parameters is carried out by (automatically) 
cycling through several TICC representations. In principle,
any (of the infinitely many possible) TICC set is allowed for which 
the $U$ transformation matrix satisfies Eqs.~(\ref{validDef1})--(\ref{validDef4}).
This multi-coordinate optimization procedure is known as multi-channel optimization
in the literature \cite{suzukivarga} where channel refers to a particular coordinate selection.

In the competitive selection procedure, to generate a new basis function, 
the basis function parameters are sampled from a normal probability distribution. 
The mean and variance values, which determine the distribution used, are determined
during the calculations by analyzing the already selected basis-set parameters.

\subsection{Numerical results}

In this section, we present numerical results of calculations carried out in laboratory-fixed Cartesian coordinates.
For the optimization of the basis function parameters we used several coordinate sets (``channels'')
in order to find more efficiently the optimal
parameter set describing the correlations (and in general, interactions) 
between pairs and groups of particles.

Tables~\ref{TAB:GeneralizedConstr1} and \ref{TAB:GeneralizedConstr2} present
numerical results of this procedure obtained 
for the ground state of the para-H$_2^+$ ($N=0$) and para-H$_2$ ($N=0$) molecular species using
the \texttt{FECG} and the \texttt{ECG-GVR}-type functions.
In the tables we show both the full LFCC energies, which include translational effects
as well as the ``corrected'', translation-free (``translationally invariant'', TI) energies, 
which are indeed smaller and which reproduce the values obtained in some translationally
invariant formulation of the Hamiltonian in the literature.

\begin{table*}[h]
\centering
\caption{\label{TAB:GeneralizedConstr1} 
  Pre-Born--Oppenheimer ground-state energies, in Hartree atomic units, of 
  the para spin state of H$_2^+=\lbrace\text{p}^+,\text{p}^+,\text{e}^-\rbrace$ as well 
  the para spin state of H$_2=\lbrace\text{p}^+,\text{p}^+,\text{e}^-,\text{e}^-\rbrace$.
  The results were obtained with the FECG-type functions, which are not angular momentum eigenfunctions, and hence
  the expectation value of the translationally invariant total orbital angular momentum squared operator, $\hat{N^2}_{\text{TI}}$, is also given.
 }
{\footnotesize 
  \begin{tabular} {@{\hspace{3mm}} c @{\hspace{5mm}} c @{\hspace{5mm}} c @{\hspace{5mm}} c @{\hspace{5mm}} c @{\hspace{5mm}} c @{\hspace{3mm}}}
  \hline
  \hline
  $\langle\hat{H}\rangle_{\text{LFCC}}$   &
  $\eta$   &
  $^a\langle\hat{H}\rangle_{\text{TI}}$   &
  $^b\langle\hat{N}^2\rangle_{\text{TI}}$  &
  $^c\eta_{\text{TI}}$     \\[1ex]
  \hline
  \multicolumn{5}{c}{ $p-$H$_2^+$ (ground state) $N_\text{b}=400$ } \\
  $-0.596231$   &  $10^{-2}$    &   $-0.597024$   &  $11.26$   &  $10^{-4}$  \\
  $-0.596988$   &  $10^{-2}$    &   $-0.597012$   &  $ 9.11$   &  $10^{-4}$  \\
  $-0.593754$   &  $10^{-2}$    &   $-0.597032$   &  $18.30$   &  $10^{-4}$  \\
  $-0.596845$   &  $10^{-2}$    &   $-0.597006$   &  $10.81$   &  $10^{-4}$  \\
  $-0.595096$   &  $10^{-2}$    &   $-0.597044$   &  $11.48$   &  $10^{-4}$  \\
  \hline
  \multicolumn{5}{c}{ $p-$H$_2$ (ground state) $N_\text{b}=600$ } \\
  $-1.162147$   &  $10^{-2}$    &  $-1.162686$  &   $10.59$    &  $10^{-4}$  \\
  $-1.162263$   &  $10^{-3}$    &  $-1.162696$  &   $10.22$    &  $10^{-4}$  \\
  $-1.161655$   &  $10^{-2}$    &  $-1.162721$  &   $14.47$    &  $10^{-4}$  \\
  $-1.161490$   &  $10^{-2}$    &  $-1.162669$  &   $15.01$    &  $10^{-4}$  \\
  $-1.160502$   &  $10^{-1}$    &  $-1.162690$  &   $19.66$    &  $10^{-4}$  \\
  \hline
  \hline
  \end{tabular}
}
\caption*{ \footnotesize{ 
    $^a$ translationally invariant energy expectation value obtained by eliminating CM \\
         contributions from the total kinetic energy; \\
    $^b$ translationally invariant total angular momentum squared expectation value $\langle\hat{N}^2\rangle_{\text{TI}}$ \\
    $^c$ translationally invariant virial coefficient, $\eta_{\text{TI}}=\left|1+\langle\Psi|\hat{V}|\Psi\rangle/2\langle\Psi|\hat{T}|\Psi\rangle_{\text{TI}}\right|$ 
} }
\end{table*}

\begin{table*}[h]
\centering
\caption{\label{TAB:GeneralizedConstr2}
  Pre-Born--Oppenheimer ground-state energies, in Hartree atomic units, of 
  the para spin state of H$_2^+=\lbrace\text{p}^+,\text{p}^+,\text{e}^-\rbrace$ as well 
  the para spin state of H$_2=\lbrace\text{p}^+,\text{p}^+,\text{e}^-,\text{e}^-\rbrace$.
  The results were obtained with \texttt{ECG-GVR}-type functions with $K_{\text{max}}=20$.
        }
{\footnotesize
  \begin{tabular} {r @{\hspace{3mm}} c @{\hspace{5mm}} c @{\hspace{5mm}} c @{\hspace{5mm}} c @{\hspace{5mm}} c @{\hspace{5mm}} c @{\hspace{5mm}} c @{\hspace{3mm}}}
  \hline
  \hline
  &
  $\langle\hat{H}\rangle_\text{LFCC}$    &  $\eta$  &
  $^a\langle\hat{H}\rangle_{\text{TI}}$  &
  $^b\eta_{\text{TI}}$   &
  $^c\delta E/\mu{\text E}_{\text h}$    \\[1ex]
  \hline
  \multicolumn{5}{c}{ $p-$H$_2^+$ ($N=0$, $M_N=0$) \hspace{2mm} $N_\text{b}=180$ \hspace{2mm} } \\
  SC$^d$  &  $-0.59(67)$   &  $10^{-3}$   &  $-0.597138979$  &  $10^{-8}$   &  $-0.084$  \\
          &  $-0.59(67)$   &  $10^{-3}$   &  $-0.597139061$  &  $10^{-8}$   &  $-0.002$  \\
          &  $-0.59(65)$   &  $10^{-3}$   &  $-0.597139059$  &  $10^{-8}$   &  $-0.004$  \\
          &  $-0.59(65)$   &  $10^{-2}$   &  $-0.597139057$  &  $10^{-8}$   &  $-0.006$  \\
          &  $-0.59(61)$   &  $10^{-2}$   &  $-0.597139059$  &  $10^{-8}$   &  $-0.004$  \\
          &  $-0.59(55)$   &  $10^{-2}$   &  $-0.597139058$  &  $10^{-8}$   &  $-0.006$  \\[1ex]
  \hline
  \multicolumn{5}{c}{ $p-$H$_2$ ($N=0$, $M_N=0$) \hspace{2mm} $N_\text{b}=500$ \hspace{2mm} } \\
  SC$^d$  &  $-1.16(35)$   &  $10^{-3}$   &  $-1.164024880$  &  $10^{-7}$   &  $-0.146$  \\
          &  $-1.16(38)$   &  $10^{-3}$   &  $-1.164025023$  &  $10^{-8}$   &  $-0.007$  \\
          &  $-1.16(36)$   &  $10^{-2}$   &  $-1.164025026$  &  $10^{-8}$   &  $-0.004$  \\
          &  $-1.16(35)$   &  $10^{-2}$   &  $-1.164025026$  &  $10^{-8}$   &  $-0.004$  \\
          &  $-1.16(30)$   &  $10^{-2}$   &  $-1.164025028$  &  $10^{-8}$   &  $-0.002$  \\
          &  $-1.16(31)$   &  $10^{-1}$   &  $-1.164025024$  &  $10^{-8}$   &  $-0.006$  \\
  \hline
  \hline
  \end{tabular}
}
\caption*{ \footnotesize{
    $^a$ translationally invariant energy expectation values obtained by eliminating CM \\
         contributions from the total kinetic energy; \\
    $^b$ translationally invariant virial $\eta_{\text{TI}}=\left|1+\langle\Psi|\hat{V}|\Psi\rangle/2\langle\Psi|\hat{T}|\Psi\rangle_{\text{TI}}\right|$ \\
    $^c$ $\delta E = E({\text{Ref.}})-\langle\hat{H}\rangle_{\text{TI}}$: \hspace{2mm} \\
         $E_{p-{\text{H}}_2^+}/{\text E}_{\text h}=-0.597139063$ from Ref.~\cite{H2+_1,H2+_2}, \hspace{2mm}
         $E_{p-{\text{H}}_2}/{\text E}_{\text h}=-1.164025030$ from Ref.~\cite{Pachucki2009,BuLeStAd09} \\
    $^d$ single-channel calculation corresponding to a single Jacobi-coordinate set.
} }
\end{table*}

The translation-free energies are obtained by using
the CM-elimination formulae in Eqs.~(\ref{ecg-transl}), (\ref{GENCONSTR:fecg}) and (\ref{GENCONSTR:ecggvr}) 
derived in the earlier sections.
In the case of \texttt{FECG}s functions, we also calculate 
translationally invariant total angular momentum squared expectation values
$\langle \hat{N}^2\rangle_{\text{TI}}$ to observe 
the contamination from excited rotational states.
The systematic study of these contributions and the analytical expressions
for this expectation value will be the presented
in a later study.

In the multi-channel optimization approach, we have included every possible set of Jacobi coordinates, 
``heavy-particle''-centered (HCP) coordinates as well as the center-of-mass-centered (CMC) coordinates are included. 
(The optimized basis function parameters are deposited in the Supplementary Material \cite{som}.)
The virial coefficient, $\eta=\big|1+\langle\Psi|\hat{V}|\Psi\rangle/2\langle\Psi|\hat{T}|\Psi\rangle_\text{TI}\big|$, vanishes for the exact 
solution (according to the virial theorem \cite{suzukivarga}), 
so it is used as an additional indicator for the overall quality of the 
variationally optimized wave function.

The FECG-type functions are not eigenfunctions of the total angular momentum operators, $\hat{N}^2$ and $\hat{N}_z$, and the parity.
These symmetry properties of the exact solution are restored numerically 
by the variational optimization procedure. In the calculations, 
the obtained total angular momentum expectation value, $\langle N^2\rangle_\text{TI}$ (see Table~\ref{TAB:GeneralizedConstr1})
is about 10, which corresponds to an effective angular momentum value of about 2.7 to be compared with the $N=0$ value of 
the absolute ground state. We include these results in the present article in order to explore the numerical behavior of FECG-type functions.
Future work might consider numerical techniques, which project the FECG functions onto irreps of the SO(3) rotation group.

In order to reproduce literature data computed in some translationally invariant representation of the coordinates and the Hamiltonian,
we also used the ECG-GVR-type functions (see Table~\ref{TAB:GeneralizedConstr2}), 
which are eigenfunctions of  the total angular momentum operators and also the space inversion.
Our results reproduce the literature data within a few nano Hartree accuracy. 
The significantly lower number of the basis functions (500 with respect to 2000) 
in comparison to earlier work using a single TICC set in the optimization \cite{Matyus2012,Matyus2013,Benjamin2013}, indicate the efficiency 
of the multi-channel optimization procedure developed in the present work. In spite of the multiple coordinate sets used for the parameter
optimization, we solve the Schr\"odinger equation (and calculate integrals)
in simple laboratory-fixed Cartesian coordinates. Translation-free energies
are obtained after the elimination of center-of-mass effects 
(compare the $\hat{H}_\text{LFCC}$ and $\hat{H}_\text{TI}$ columns of Table~~\ref{TAB:GeneralizedConstr2}).

\section{Summary and conclusions}

The article presents further progress about the solution of the many-particle Schr\"odinger equation 
in laboratory-fixed Cartesian coordinates (LFCCs). We extend our earlier work using 
explicitly correlated Gaussian (ECG) functions \cite{Benjamin2013} to be applicable with
a more efficient basis-function parameter optimization procedure, called multi-channel optimization.
Multi-channel optimization relies on the optimization of the interaction of 
several possible pairs and groups of particles (``channels'') of the many-particle system by repeatedly
changing the pairing or grouping of the particles. This idea is realized in our work by transforming 
the basis function parameterization back and forth during the optimization procedure between the different 
particle groups or channels, which, after all, are represented by some coordinate set, 
while we solve the Schr\"odinger equation (Hamiltonian, matrix elements, etc.) 
in simple LFCCs and obtain translationally invariant (TI) properties.

In order to implement these general ideas in an algorithm and computer code, 
we study the form of the basis functions and the mathematical expressions of 
the Hamiltonian matrix elements upon the transformation of the coordinates between LFCCs and 
various translationally invariant sets of Cartesian coordinates and the center-of-mass coordinates 
(TICCs and CMCCs). We also work out the formal equations  
which prove that the various (infinitely many) possible sets of TICCs form a closed set
and can be combined arbitrarily without introducing any contamination from the center of mass coordinates
(see Section~\ref{ch:fundam}).
Using these results, we identify the center-of-mass (CM) terms in the kinetic-energy integral 
expressions for three particular types of ECG functions, 
which is necessary for the multi-channel implementation. 
Translationally invariant energies are obtained from an LFCC Hamiltonian by eliminating
these CM terms during the course of the integral evaluation procedure, 
performed in LFCCs.

The applicability and efficiency of this new algorithm and computer code is demonstrated for the
ground state of the three-particle H$_2^+=\lbrace\text{p}^+,\text{p}^+,\text{e}^-\rbrace$
as well as of the four-particle H$_2=\lbrace\text{p}^+,\text{p}^+,\text{e}^-,\text{e}^-\rbrace$ molecular systems.
We solve the many-particle Schr\"odinger equation in laboratory-fixed Cartesian coordinates
and eliminate the translational contamination during the integral evaluation, while we
optimize the basis-function parameters using multiple channels (coordinates) including 
all possible Jacobi 
coordinates, all possible heavy-particle-centered coordinate arrangements, as well as the 
center-of-mass-centered coordinate set.

Our present LFCC formalism allows an increased flexibility of the basis functions and a better energy convergence.
It is an alternative to the traditional approaches using some set of TICCs with the Cartesian coordinates
of the center of mass explicitly separated out from the Hamiltonian. 
%The ideas and formalism developed
%in this work are transferable to the Dirac theories and will allow us to calculate translationally invariant
%energies from the (simple and standard LFCC) Dirac Hamiltonian(s).

\section*{Acknowledgments}

This work was supported by ETH Zurich and by the Schweizerischer Nationalfonds (Project No. SNF $200020\_169120$).
EM acknowledges financial support from a PROMYS Grant (no. IZ11Z0\_166525)  
of the Swiss National Science Foundation and the the COST Action MOLIM (CM1405).

\appendix
%--------------------------------------------------------------------------------------------------------------------------------
\section{Center-of-mass contributions to the overlap integral for \texttt{pECG} functions}  \label{APP:cA_det_terms}

The normalized overlap matrix element $IJ$-th for \texttt{pECG} functions is
\begin{equation}
\frac{\left\langle\phi_I|\phi_J\right\rangle}{\left(\left\langle\phi_I|\phi_I\right\rangle\left\langle\phi_J|\phi_J\right\rangle\right)^{\frac{1}{2}}} =
\left(\frac{\left|2\bar{A}_I^{(r)}\right|^{\frac{1}{2}}\left|2\bar{A}_J^{(r)}\right|^{\frac{1}{2}}}
{\left|\bar{A}_I^{(r)}+\bar{A}_J^{(r)}\right|}\right)^{\frac{3}{2}}.
\end{equation}

Similarly to Sec.~(\ref{SEC:constraints}) we identify $c_A$-related terms, which are associated to the 
center-of-mass coordinate. 
Firstly, we can rewrite the expressions as
\begin{equation}
\left( \left|2U_x^{\text{T}}\bar{A}_I^{(x)}U_x\right| \cdot \left|2U_y^{\text{T}}\bar{A}_J^{(y)}U_y\right| \cdot
\left|\bar{A}_{IJ}^{(r)}\right| \cdot \left|\bar{A}_{IJ}^{(r)}\right| \right)^{\frac{3}{4}} ,
\end{equation}
and employ the properties of determinants, $\left|A\cdot B\right|=\left|A\right|\cdot\left|B\right|=\left|B\cdot A\right|$ and 
$\left|A^{-1}\right|=\left|A\right|^{-1}$, to arrive at
\begin{equation}
\left| 4\cdot U_x^{\text{T}}\bar{A}_{IJ}^{(r)}U_y \cdot \bar{A}_I^{(x)} \cdot U_y^{\text{T}}\bar{A}_{IJ}^{(r)}U_x \cdot \bar{A}_J^{(y)} \right| .
\end{equation}
If different $c_{A_I}$ and $c_{A_J}$ values were allowed 
for the $I$th and $J$th basis functions, we obtained
\begin{equation}
\left| 4 \cdot 
\left( \begin{array}{cc} \mathcal{A}_{IJ}^{-1} & 0 \\ 0 & \frac{1}{c_{A_I}+c_{A_J}} \end{array} \right)
\left( \begin{array}{cc} \mathcal{A}_I^{(x)} & 0 \\ 0 & c_{A_I} \end{array} \right)
\left( \begin{array}{cc} \mathcal{A}_{IJ}^{-1} & 0 \\ 0 & \frac{1}{c_{A_I}+c_{A_J}} \end{array} \right)
\left( \begin{array}{cc} \mathcal{A}_J^{(y)} & 0 \\ 0 & c_{A_J} \end{array} \right) \right|.
\end{equation}
Contributions from the $c_A$ factors cancel only if $c_{A_I} = c_{A_J}$.

%\bibliographystyle{jcp}
%\bibliography{genconstr}
\newcommand{\Aa}[0]{Aa}

\end{document}